\documentclass[10pt]{article}
\usepackage[cp1251]{inputenc}
\usepackage[russian]{babel}
\usepackage[dvips]{epsfig}   \usepackage[dvips]{changebar}
\usepackage[dvips]{graphicx}

\def\lsim{\
  \lower-1.2pt\vbox{\hbox{\rlap{$<$}\lower5pt\vbox{\hbox{$\sim$}}}}\ }
\def\gsim{\
  \lower-1.2pt\vbox{\hbox{\rlap{$>$}\lower5pt\vbox{\hbox{$\sim$}}}}\ }
   
\usepackage{amsmath,enumerate, amsfonts, amssymb,amsthm}

\begin{document}
\title{Structure of a condensate for the Bose fluid in a vessel}
 \author{Max Tomchenko
\bigskip \\ {\small Bogolyubov Institute for Theoretical Physics} \\
 {\small 14b, Metrolohichna Str., Kiev 03680, Ukraine} \\
 {\small E-mail:mtomchenko@bitp.kiev.ua}}
 \date{\empty}
 \maketitle
 \large
 \sloppy
 \textit{We show that the ground state of the Bose fluid in a vessel-parallelepiped
can be characterized not only by
the quantum numbers  $\textbf{p}=(\pm 2\pi l_{x}/L_{x}, \pm 2\pi l_{y}/L_{y}, \pm 2\pi l_{z}/L_{z})$
($l_{x}, l_{y}, l_{z} =  1, 2,  3, \ldots$) of an expansion in the eigenfunctions $e^{i\textbf{p}\textbf{r}}$ of a free particle, but also
by  the quantum numbers $\textbf{k}_{\textbf{l}}=(\pi l_{x}/L_{x}, \pi l_{y}/L_{y}, \pi l_{z}/L_{z})$
of an expansion in the eigenfunctions  $\sin{(k_{l_{x}}x)}\sin{(k_{l_{y}}y)}\sin{(k_{l_{z}}z)}$
of a particle in the box.
In the latter case,  the one-particle condensate ``is dispersed'' over many lower levels with odd $l_{x}, l_{y}, l_{z}$.
$53\,\%$ atoms of the condensate occupy the lowest level with $\textbf{k}=(\pi/L_{x}, \pi/L_{y}, \pi/L_{z})$.
The sum of condensates on all levels is determined by the same formula $N_{c}/N=\rho_{\infty}$,
as that for the condensate on the level with $p=0$.
The proposed $\textbf{k}$-representation supplements the traditional one (with the condensate on the level $\epsilon = (\hbar p)^{2}/2m=0$)
 and allows one to consider the microstructure of a system from another side.}
\maketitle

\section{Introduction}
The problem of a condensate in the Bose fluid attracts attention
as before. The condensate was predicted theoretically
\cite{einstein25}--\cite{penronz1956} and was then discovered experimentally.
Moreover, the experimental values \cite{blag,cond2000} agree
sufficiently well with theoretical ones
\cite{penronz1956,gfmc1981}--\cite{rota2012}.
Intensively are studied the condensates in dilute  gases in
traps.
However, we meet the difficulty, which remains usually unnoticeable. A condensate in the Bose
fluid is considered as the ``macroscopic number of atoms with zero
momentum.'' But the fluid is placed  always in a vessel, and
the uncertainty relation $\hbar\triangle p_{x} \triangle x \geq
\hbar/2$ implies that the momentum of an atom cannot be zero due
to the uncertainty $\triangle p_{x} \geq 1/2L$. This point is
obviously clear: the momentum is defined for translation-invariant systems,
whereas the walls break the translational
invariance, and the momentum of an atom is not conserved due to
the collisions with walls.
 In addition, the momentum takes the discrete values
$\textbf{p}=  2\pi( j_{x}/L_{x}, j_{y}/L_{y}, j_{z}/L_{z})$ in the standard approach,
and its uncertainty $\geq 1/2L$ is comparable with the distance between adjacent values. Therefore, the collection of discrete levels
is transformed in a collection of bands or even in a single band.
 This means that the momentum is a ``bad'' quantum number. We can avoid these difficulties
through consideration $\textbf{p}$ as an eigenvalue of the Hamiltonian
only ($\epsilon = (\hbar p)^{2}/2m$). In this case, we can use the
traditional formalism and obtain, as usual, the single
macroscopically filled level with $\textbf{p}=0$; but
$\textbf{p}$ characterizes here the energy of a particle, rather than
its momentum. However, it is possible also to construct a nontraditional
approach to the description of the microstructure of systems.
Such new approach will be considered in what follows.

\section{Structure of a condensate for the Bose fluid in a vessel}
Consider the Bose fluid in a vessel at $T=0$. In the standard
approach with periodic boundary conditions (BCs), the condensate
$n_{0}=N_{\textbf{p}=0}/N$ is determined from the formula
\begin{equation}
\frac{N_{\textbf{p}}}{N}=\int d\textbf{r}_{1} d\textbf{r}_{2}
\rho(\textbf{r}_{1},\textbf{r}_{2})\frac{e^{-i\textbf{p}(\textbf{r}_{1}-\textbf{r}_{2})}}{V^2},
\label{1}  \end{equation} where $N$ is the total number of
identical Bose particles, $V$ is the volume of the system,
$\rho(\textbf{r}_{1},\textbf{r}_{2})$ is the one-particle density
matrix,
\begin{equation}
\rho(\textbf{r}_{a},\textbf{r}_{b}) = V\int d\textbf{r}_{2}\ldots d\textbf{r}_{N}
\Psi_{0}^{*}(\textbf{r}_{a},\textbf{r}_{2},\ldots,\textbf{r}_{N})\Psi_{0}(\textbf{r}_{b},\textbf{r}_{2},\ldots,\textbf{r}_{N}),
\label{2}  \end{equation}
and $\Psi_{0}$ is the wave function of the ground state of the system.
For an infinite system, we have
$\rho(\textbf{r}_{1},\textbf{r}_{2})=\rho(|\textbf{r}_{1}-\textbf{r}_{2}|)$ due to the translational invariance and the isotropy. Therefore,
\begin{equation}
n_{0}=\frac{N_{\textbf{p}=0}}{N}=\frac{1}{V^2}\int d\textbf{r}_{1}
d\textbf{r}_{2}
\rho(|\textbf{r}_{1}-\textbf{r}_{2}|) \approx \rho(|\textbf{r}_{1}-\textbf{r}_{2}|\rightarrow
\infty)\equiv \rho_{\infty}. \label{3}  \end{equation} The
presented definition of the condensate follows from the expansion of a one-particle wave function
\begin{equation}
 \Psi(\textbf{r})=\int d\textbf{p} a(\textbf{p})e^{i\textbf{p}\textbf{r}}(2\pi)^{-3}
\label{fur}  \end{equation}
in the eigenfunctions $e^{i\textbf{p}\textbf{r}}$ of the momentum operator,
which form the complete set of orthonormalized basis functions of the continuous spectrum. The exponential function
$e^{i\textbf{p}\textbf{r}}$ is an eigenfunction of the Hamiltonian and the momentum operator for a free particle
in the infinite or finite closed volume. The closed ring systems are described by periodic
BCs and are realized in the nature only in the one- or two-dimensional case.

Consider the finite-size systems with nonperiodic BCs. In
particular, the boundary conditions for the three-dimensional
systems are always nonperiodic. As was noted in \cite{gir1965},
the expansion for the system in a box should be performed
in  an eigenfunctions of this system.
Girardeau \cite{gir1965} has found the solution for $N$ point Bose
particles in a box-cube $L\times L\times L$ characterized by an
interatomic potential in the form of the $\delta$-function.  The
BCs were taken to be $\textbf{n}\nabla \psi=0$. It was found that
all atoms are on the lowest level at a repulsive potential. At an
attractive potential, the atoms are distributed over ($\sim
N^{2/5}$) lower levels, but with the microscopic ($\sim N^{3/5}$)
population of each level. In this case, the condensate exists in
the ``generalized'' sense \cite{gir1965}: the total number of
atoms on those levels turns out macroscopic.

But the real atoms are not point-like and possess a smoother
potential, than the $\delta$-function. Therefore, the wave
functions may have a different structure \cite{zero-liquid} as
compared with that of the solutions  in \cite{gir1965}. We may
expect also a different solution for the condensate. We now
consider such realistic system of Bose particles in the box
$L_{x}\times L_{y}\times L_{z}$. For simplicity, let us consider
firstly the 1D case. We model the potential of a wall as a step
with finite height:
\begin{equation}
 U_{w}(x) =
\left [ \begin{array}{ccc}
    U>0  & \   x\leq 0, \ x\geq L  & \\
    0  & \ 0<x<L. &
\label{4} \end{array} \right. \end{equation}
Let the potentials of both walls be identical (their distinction will not lead to basic changes).
The eigenfunctions of a free particle in such a box are known \cite{land3}:
 \begin{equation}
 \Psi_{l}(x) =
\left [ \begin{array}{ccc}
    c_{l}\sin{(\delta_{l})}e^{\kappa_{l}x},  & \   x\leq 0  & \\
    c_{l}\sin{(k_{l}x+\delta_{l})},  & \ 0<x<L_{x} & \label{5} \\
    c_{l}(-1)^{l+1}\sin{(\delta_{l})}e^{-\kappa_{l}(x-L)},  & \   x\geq L_{x},  &
 \end{array} \right. \end{equation}
\begin{equation}
k_{l}=\sqrt{2mE_{l}}/\hbar = \pi l/L_{x} - 2\delta_{l} /L_{x},
     \label{6a} \end{equation}
\begin{equation}
\kappa_{l}=\sqrt{2m(U-E_{l})}/\hbar >0, \ \delta_{l}=\arcsin{(\gamma_{x}k_{l}L_{x}/2)},
     \label{6b} \end{equation}
\begin{equation}
c_{l}=\left (\frac{(\sin\delta_{l})^{2}}{\kappa_{l}}+\frac{L_{x}}{2}+\frac{\sin(2\delta_{l})}{2k_{l}}\right )^{-1/2}
\approx \sqrt{\frac{2}{L_{x}}}.
     \label{6c} \end{equation}
Here, $l=1, 2, 3, \ldots$, $\gamma_{x}=\frac{\hbar}{L_{x}}\sqrt{\frac{2}{mU}}$, and the values of $\arcsin$ are taken between $0$ and $\pi/2$.
The boundary conditions involve the continuity of $\Psi_{l}(x)$, $\Psi^{'}_{l}(x)$ and are more physical
than the BCs $\textbf{n}\nabla \psi=0$ in \cite{gir1965}.
For great systems, $\gamma_{x}\ll 1$: for He$^4$ atoms at $U=100\,$K and $L_{x}=10\,$cm, we have
$\gamma_{x}\approx 4.9\times 10^{-10}$.  For  $l\ll 1/\pi\gamma_{x},$ we obtain
\begin{equation}
k_{l}\approx lk_{1}=l(1-\gamma_{x})\pi /L_{x}, \quad \delta_{l}\approx l\delta_{1}=l\pi \gamma_{x}/2.
     \label{7} \end{equation}
We note that the results of the work
are independent of the choice of a finite or infinite barrier $U$.
Indeed, for any reasonable $U\gsim 1\,$K, we have $\delta_{1}
\lsim 10^{-8}$; for not too large $l,$ all $\delta_{l}$ (\ref{7})
are small, so that we can take $\cos{\delta_{l}}\rightarrow 1$ in
formula (\ref{14}). Therefore, the final formulas (\ref{16}) and
(\ref{17}) are the same for different $U$.  But since the real
barriers are finite, we will use the formulas for finite  $U$. We
can pass to infinite $U$, by setting $\delta_{l} \rightarrow 0$ in the formulas.

Solutions (\ref{5}) describe bound states with discrete spectrum. Here, we omit the solutions $\Psi_{f}(x)$ with continuous spectrum, which
correspond to the infinite motion with $E>U.$
The solutions for the continuous and discrete spectra
are eigenfunctions of the Hermitian operator (Hamiltonian). By the theorems of quantum mechanics,
these functions form the complete orthonormalized set of basis functions, in which any function
$\Psi(x)$ with the appropriate smooth properties and given on the interval $x \in ]-\infty, \infty [$ can be expanded.
Therefore, any solution for a free particle localized in a 1D box (\ref{4})
can be written in the form
\begin{equation}
\Psi(x)=\sum\limits_{l}b_{l}\Psi_{l}(x), \  \
b_{l}=\int\limits_{-\infty}^{\infty}dx \Psi(x)\Psi_{l}^{*}(x),
     \label{9} \end{equation}
and the probability for the particle in the state $\Psi(x)$ to have the wave number $k_{l}$ is equal to $|b_{l}|^{2}$.

Let us consider the Bose fluid in a vessel.
In the presence of a vessel, the momentum is not a quite good quantum number. Therefore,
we are not based on the momentum operator, while choosing the eigenvalues for the expansion of the density matrix. Then the exponents
$e^{ip_{x}x}$ lose the status of exceptional functions: now, we can use only the eigenfunctions of the Hamiltonian
of a free particle,
which are the collections of all possible linear combinations of the form
$ae^{ip_{x}x}+be^{-ip_{x}x}$ with different $p_{x}$. Among those collections, we mark out two ones,
which are the complete sets of the basis functions:
the exponents $e^{ip_{x}x}$, where $p_{x}$ are multiple to $2\pi/L$ (they give solutions for the quasiparticles in the form of traveling waves, and the phonons in real systems
are namely traveling wave packets, rather than standing waves; we call it $\textbf{p}$-representation), and the sines $\sin{(k_{l}x+\delta_{l})}$ (\ref{5})
(since they take the boundary conditions into account, by ensuring the decrease of the wave functions near the boundary;
we call it the $\textbf{k}$-representation). Here,  the wave vectors $\textbf{k}$ and $\textbf{p}$ are similar to the momentum,
but they are not the momentum, but a quantum numbers of the Hamiltonian.
None of these sets of eigenfunctions can be called the best. In our view, the exponential functions describe better the quasiparticles,
whereas the sines do the ground state and the condensate.

In the $\textbf{p}$-representation, we obtain obviously the well-known results with the condensate on a level with $\textbf{p}=0$.
Of a significant interest is the $\textbf{k}$-representation based on the functions $\Psi_{l}(x)$ (\ref{5}).
We now consider it in more details.
The wave vector takes the values
\begin{equation}
\textbf{k}=\textbf{k}_{\textbf{l}}=  \left (\frac{\pi l_{x}-2\delta_{l_{x}}}{L_{x}},
\frac{\pi l_{y}-2\delta_{l_{y}}}{L_{y}},
\frac{\pi l_{z}-2\delta_{l_{z}}}{L_{z}}\right )
     \label{12} \end{equation}
with $\ l_{x}, l_{y}, l_{z} =  1, 2,3, \ldots$
We omit the negative components, since the wave function of a particle for them differs only by a sign (these are equivalent states).
For the Bose fluid states of the discrete spectrum, the number of particles with the wave vector $\textbf{k}$ is
\begin{equation}
\frac{N_{\textbf{k}}}{N}=\frac{1}{V}\int\limits_{-\infty}^{\infty} d\textbf{r}_{1} d\textbf{r}_{2}
\rho(\textbf{r}_{1},\textbf{r}_{2})\Psi^{*}_{l_{x}}(x_{1})\Psi_{l_{x}}(x_{2})
 \Psi^{*}_{l_{y}}(y_{1})\Psi_{l_{y}}(y_{2})\Psi^{*}_{l_{z}}(z_{1})\Psi_{l_{z}}(z_{2}).
\label{10a}  \end{equation}
Let us verify the correctness of the coefficient $V^{-1}$. The property of eigenfunctions
\begin{equation}
\sum\limits_{\textbf{l}}\Psi^{*}_{\textbf{l}}(\textbf{r}_{1})\Psi_{\textbf{l}}(\textbf{r}_{2})+
 \int d\textbf{f}\Psi^{*}_{\textbf{f}}(\textbf{r}_{1})\Psi_{\textbf{f}}(\textbf{r}_{2}) = \delta(\textbf{r}_{1}-\textbf{r}_{2})
     \label{ortog2} \end{equation}
and the normalization of $\Psi_{0}$ yield
\begin{equation}
\sum\limits_{\textbf{k}}\frac{N_{\textbf{k}}}{N}+  \int \textbf{df}\frac{N_{\textbf{f}}}{N} = 1,
     \label{norm} \end{equation}
i.e., the coefficient is proper.
The integral over the continuous spectrum is probably much less than 1.

Let us determine $N_{\textbf{k}}$. According to (\ref{10a}), we need to integrate over the whole volume, inside
and outside of the box. But we do not know the solution for $\Psi_{0}$
outside of the box. However, it is clear that this solution decreases exponentially,  with the penetration distance $ \sim 1\,\mbox{\AA}$
for realistic walls. Therefore, we neglect the region outside of the box. Then, in the 3D case, we obtain
\begin{eqnarray}
&\frac{N_{\textbf{k}_{\textbf{l}}}}{N}&=\frac{1}{V}\int\limits_{0}^{L_{x}, L_{y}, L_{z}} d\textbf{r}_{1} d\textbf{r}_{2}
\rho(\textbf{r}_{1},\textbf{r}_{2})c^{2}_{l_{x}}c^{2}_{l_{y}}c^{2}_{l_{z}}
\sin{(k_{l_{x}}x_{1}+\delta_{l_{x}})}\sin{(k_{l_{x}}x_{2}+\delta_{l_{x}})} \nonumber \\ &\times &
\sin{(k_{l_{y}}y_{1}+\delta_{l_{y}})}
\sin{(k_{l_{y}}y_{2}+\delta_{l_{y}})}
\sin{(k_{l_{z}}z_{1}+\delta_{l_{z}})}
\sin{(k_{l_{z}}z_{2}+\delta_{l_{z}})}.
\label{10b}  \end{eqnarray}
We now have $\rho(\textbf{r}_{1},\textbf{r}_{2})\neq \rho(\textbf{r}_{1}-\textbf{r}_{2})$, generally speaking. At
distances from the wall much larger than the interatomic distance $\bar{R}$, the properties of $\Psi_{0}$ are the same \cite{zero-liquid}
as those for an infinite system. Here, we consider only the general properties of
$\Psi_{0}$ (there is no exact coincidence for $\Psi_{0}$, see formula (\ref{19}) below).
In particular, at $\textbf{r}_{1}$ and $\textbf{r}_{2}$ far from the walls and $|\textbf{r}_{1}-\textbf{r}_{2}|\gg \bar{R},$
the relation $\rho(\textbf{r}_{1},\textbf{r}_{2})=\rho_{\infty}=const$ is true. With regard for it, relation (\ref{10b}) yields
\begin{equation}
\frac{N_{\textbf{k}_{\textbf{l}}}}{N}=\left(
\frac{8c_{l_{x}}c_{l_{y}}c_{l_{z}}\cos{\delta_{l_{x}}}\cos{\delta_{l_{y}}}
\cos{\delta_{l_{z}}}}{k_{l_{x}}k_{l_{y}}k_{l_{z}}}\right )^2 \frac{\rho_{\infty}}{V},
\label{14}  \end{equation}
if $l_{x}, l_{y}, l_{z}$ are odd, and $N_{\textbf{k}_{\textbf{l}}}/N=0$,
if at least one of $l_{x}, l_{y}, l_{z}$ is even.
Let us consider odd $l_{x}, l_{y}, l_{z}$.
The value of $N_{\textbf{k}_{\textbf{l}}}$ is large at small $\textbf{k}_{\textbf{l}}$, when relations
(\ref{7}) hold. In this case,
\begin{equation}
\frac{N_{\textbf{k}_{\textbf{l}}}}{N} \approx
\frac{8^{3}\rho_{\infty}}{\pi^{6}l_{x}^{2}l_{y}^{2}l_{z}^{2}}.
\label{16}  \end{equation}
Since $\rho_{\infty}\sim 1$, we obtain the condensate $N_{\textbf{k}_{\textbf{l}}}\sim N$ for all levels with small $l_{x}, l_{y}, l_{z}$.
The sum of condensates on all levels
\begin{equation}
n_{c}=\sum\limits_{\textbf{k}}\frac{N_{\textbf{k}}}{N} \approx
\frac{8^{3}\rho_{\infty}}{\pi^{6}}
\sum\limits_{j_{x}, j_{y}, j_{z}}\frac{1}{(1+2j_{x})^{2}(1+2j_{y})^{2}(1+2j_{z})^{2}}
=\rho_{\infty}
\label{17}  \end{equation}
($j_{x}, j_{y}, j_{z}=0,  1,  2,  3, \ldots$) is exactly equal to the condensate on the level $\textbf{p}=0$ for the $\textbf{p}$-representation.
This interesting property is general. We denote $V^{-1/2}= |\textbf{r}\rangle $,
$\rho_{\infty}=\hat{\rho}$, $\Psi_{l_{x}}(x)\Psi_{l_{y}}(y)\Psi_{l_{z}}(z)=|\textbf{l}\rangle$ and
expand the function $\Psi(\textbf{r})$, equal to $V^{-1/2}$ in the box and to zero outside of it, in the functions $|\textbf{l}\rangle$
and the continuous-spectrum functions of a particle in the box. The latter enter the expansion with zero coefficients. Therefore,
$\Psi(\textbf{r})=\sum\limits_{\textbf{l}}c_{\textbf{l}}|\textbf{l}\rangle$,
$c_{\textbf{l}}=\langle \textbf{r}|\textbf{l} \rangle$. Then,
in the approximation $\rho(\textbf{r}_{1},\textbf{r}_{2})=\rho_{\infty},$ we have
\begin{eqnarray}
&&\sum\limits_{\textbf{p}}N_{\textbf{p}}/N = N_{\textbf{p}= 0}/N  =\rho_{\infty} =\int d\textbf{r}\rho_{\infty} (e^{i \textbf{pr}}/V)|_{\textbf{p}\rightarrow 0}
\nonumber \\ & = & \langle \textbf{r}| \hat{\rho}|\textbf{r} \rangle
= \langle \textbf{r}|\textbf{l} \rangle \langle \textbf{l}| \hat{\rho}|\textbf{m} \rangle \langle \textbf{m}|\textbf{r} \rangle
= \sum\limits_{\textbf{l},\textbf{m}}c_{\textbf{l}}c^{*}_{\textbf{m}}\langle \textbf{l}| \hat{\rho}|\textbf{m} \rangle \nonumber \\
&=& \sum\limits_{\textbf{l}}|c_{\textbf{l}}|^{2}\langle \textbf{l}| \hat{\rho}|\textbf{l} \rangle =
\rho_{\infty}\sum\limits_{\textbf{l}}|c_{\textbf{l}}|^{2}= \sum\limits_{\textbf{k}_{\textbf{l}}}N_{\textbf{k}_{\textbf{l}}}/N.
\label{18}  \end{eqnarray}
We do not use the specific form of the functions $|\textbf{l}\rangle.$ Therefore, the summary condensate
$\sum\limits_{\textbf{k}_{\textbf{l}}}N_{\textbf{k}_{\textbf{l}}}/N$
is the same for any basis of eigenfunctions.
This property is related to the  approximation $\rho(\textbf{r}_{1},\textbf{r}_{2})=\rho_{\infty},$ for which
$\langle \textbf{l}| \hat{\rho}|\textbf{m} \rangle= \rho_{\infty}\delta_{\textbf{l},\textbf{m}}$,
 and to the fact that $\rho(\textbf{r}_{1},\textbf{r}_{2})\approx \rho_{\infty}=const$ can be expanded in various bases.

The real value of $n_{c}$ is somewhat less than that in (\ref{17}). The condensate corresponds to the macroscopic filling of a level,
but the sum (\ref{17}) includes also large values of $j_{x}, j_{y}, j_{z}$, at which the filling is small.
We call the level macroscopically filled, if $N_{\textbf{k}}> q_{c}N$.
The value of $q_{c}$ is conditional, but the estimate $q_{c}\sim (1-100)\times N^{-1/3}$ seems reasonable. The numerical analysis indicates that,
at $q_{c}=0.001,$ we have $n_{c}\approx 0.914\rho_{\infty}$ and the number of condensate levels $N_{c}= 49$.
At $q_{c}=10^{-6},$ we obtain $n_{c}\approx 0.994\rho_{\infty}$ and $N_{c}=3920$. Whereas
$n_{c}\approx 0.999\rho_{\infty}$ and $N_{c}\approx 61400$ at $q_{c}=10^{-8}$.
For great systems $N^{1/3}\gsim 10^8.$ In this case $q_{c}\sim 10^{-6}$-$10^{-8},$ and $n_{c}$
differs slightly from $n_{c}=\rho_{\infty}$ (\ref{17}).

It is clear that, for the 1D and 2D cases, the properties of the condensate are analogous for $T=0$.

For a different shape of the vessel, we need to use different basis functions at the expansion,
but the summary condensate is not changed, according to (\ref{18}). If we consider the realistic potential of a wall
(smoother than that of a step and like the interaction potential of two helium atoms), the summary condensate is not changed
as well, but, probably,
will be more strongly dispersed over levels.

The additional information is given by the exact wave function
$\Psi_{0}$ \cite{zero-liquid} of a system of $N$ interacting Bose
particles in a box-parallelepiped with impermeable walls
($U=\infty$):
\begin{equation}
\Psi_{0}=\Psi^{b}_{0}e^{S^{(1)}_{w}}\prod\limits_{j}\left \{ \sin{(k_{1_{x}}x_{j})}\sin{(k_{1_{y}}y_{j})}\sin{(k_{1_{z}}z_{j})}\right \},
\label{19}  \end{equation}
\begin{equation}
 S_{w}^{(1)} = \sum\limits_{\textbf{q}\neq 0}S_{1}^{(1)}(\textbf{q})\rho_{-\textbf{q}} +
 \sum\limits_{\textbf{q},\textbf{q}_{1}\neq 0}^{\textbf{q}+\textbf{q}_{1}\neq 0}
\frac{S_{2}^{(1)}(\textbf{q},\textbf{q}_{1})}{\sqrt{N}}\rho_{\textbf{q}_{1}}\rho_{-\textbf{q}_{1}-\textbf{q}}+\ldots,
     \label{20} \end{equation}
where $\textbf{k}_{1} =(k_{1_{x}},k_{1_{y}},k_{1_{z}})= (\pi /L_{x}, \pi /L_{y}, \pi /L_{z})$,
$\rho_{\textbf{k}} = \frac{1}{\sqrt{N}}
   \sum\limits_{j=1}^{N}e^{-i\textbf{k}\textbf{r}_j}$,
 and $\Psi^{b}_{0}$ is the bulk part of $\Psi_{0}$:
\begin{equation}
\ln{\Psi_{0}^{b}}=
 \sum\limits_{\textbf{q}_{1}\neq 0}\frac{a_{2}(\textbf{q}_{1})}{2!}\rho_{\textbf{q}_{1}}
   \rho_{-\textbf{q}_{1}}+
 \sum\limits_{\textbf{q}_{1},\textbf{q}_{2}\neq 0}^{\textbf{q}_{1}+\textbf{q}_{2}\not= 0}
  \frac{a_{3}(\textbf{q}_{1},\textbf{q}_{2})}{3!\sqrt{N}}
 \rho_{\textbf{q}_{1}}\rho_{\textbf{q}_{2}}\rho_{-\textbf{q}_{1} - \textbf{q}_{2}}
  + \ldots
\label{21}  \end{equation}
The components of $\textbf{q}$  are multiple to $2\pi/L$, and the components of  $\textbf{q}_{j}$ are multiple to $\pi/L$.
If we expand both exponential functions in (\ref{19}) in a series, represent $\rho_{\textbf{k}}$ in terms of cosines and sines,
and convolve them with sines from the bare product
of sines in (\ref{19}), we obtain the sum of products of $N$ sines and cosines  with wave vectors
multiple to $\pi/L$. This yields two conclusions for particles in a box.  1) The wave vectors of interacting
particles ``are quantized'' in the same way like those of free particles.
2) If the interatomic interaction is switched-off,    $\Psi_{0}$ is reduced to a product of sines
(since $S_{w}^{(1)}\rightarrow 0$, $\Psi^{\infty}_{0}\rightarrow 1$ \cite{zero-liquid}), i.e., to
the solution for the ground state of $N$ free particles, when all particles are in the condensate with $\textbf{k}=\textbf{k}_{1}$.
The structure of condensate (\ref{17}) agrees with this solution: if the interaction tends to zero,
the atoms from above-condensate levels ``fall'' on the condensate levels; if the interaction
is switched-off completely, the condensate levels become empty by jump: the atoms transit to the lowest level with $\textbf{k}=\textbf{k}_{1}$.
But if the condensate is defined in the standard way (\ref{1}), we cannot obtain $N$ particles on the
level with $\textbf{k}=\textbf{k}_{1}$ at the switching-off of the interaction. In this relation, the new representation (\ref{17}) of a condensate is better.
The condensate measured in experiment is, in fact, the quantity $\rho_{\infty}$.

Let us compare the formulas for the number of above-condensate atoms, $N_{\textbf{p}}$,
in the $\textbf{k}$- and $\textbf{p}$-representations for $k, p \gg \pi/L$.
Let us expand the density matrix in a Fourier series:
\begin{equation}
 \rho(\textbf{r}_{1}-\textbf{r}_{2}) = \frac{1}{V^{*}}
 \sum\limits_{\textbf{q}}f(\textbf{q})e^{i\textbf{q}(\textbf{r}_{1}-\textbf{r}_{2})}.
     \label{30} \end{equation}
For the $\textbf{p}$-representation, we use periodic BCs. Then $V^{*}=V,$ and $\textbf{q}$ runs the values
$2\pi(j_{x}/L_{x}, j_{y}/L_{y}, j_{z}/L_{z}).$ Relation (\ref{1}) yields
\begin{equation}
N_{\textbf{p}}=f(\textbf{p})N/V.
\label{31}  \end{equation}
For the $\textbf{k}$-representation, $V^{*}= 8V$, and $\textbf{q}$ runs the values
$\pi(j_{x}/L_{x}, j_{y}/L_{y}, j_{z}/L_{z})$, and relations (\ref{10b}) and (\ref{30}) yield
\begin{equation}
N_{\textbf{k}}=r_{\textbf{l}}f(\textbf{k})N/V, \quad  r_{\textbf{l}}= r_{l_{x}}r_{l_{y}}r_{l_{z}},
\label{32}  \end{equation}
\begin{equation}
r_{l_{x}}=\frac{1}{6}+\frac{\cos^{2}{\delta_{l_{x}}}}{3}  + \frac{\sin^{2}{\delta_{l_{x}}}}{2\delta_{l_{x}}^2},
\label{33}  \end{equation}
$l_{x}=1, 2, \ldots, l_{x}^{max}-1$. The value of $l_{x}^{max}= 1/\delta_{1}$ can be determined from the condition
$\delta_{l_{x}^{max}}=\pi/2$ and gives the maximum value of $k_{l_{x}}$.
The analogous result holds for the $y$- and $z$-components. The vectors $\textbf{k}> \textbf{k}^{max}$ correspond to the continuous spectrum.
At $\textbf{k}\leq \textbf{k}^{max}/2$ (for all components), $ r_{\textbf{l}} \approx 1$  is true.
At larger $\textbf{k},$
$ r_{\textbf{l}}$ decreases, as $\textbf{k}$ increases, and $ r_{\textbf{l}}(\textbf{k}^{max}) \approx 0.05$.
For He$^4$ atoms at $U=25\,$K, we have $k^{max}=\sqrt{6mU}/\hbar \approx 7.2\,\mbox{\AA}^{-1}$,
which equals to the  threshold momentum $k^{m}\approx 3.6\,\mbox{\AA}^{-1}$ of the dispersion curve of He II.
We do not exclude that this property explain the  threshold.
Thus, $ r_{\textbf{l}} \approx 1$ for all $\textbf{k}$ for which $f(\textbf{k})$ is not too small,
and the formulas for $N_{\textbf{k}}$ and $N_{\textbf{p}}$ are identical.
In view of the positiveness of the components of $\textbf{k}$,
we have $\sum\limits_{\textbf{k}}= \sum\limits_{\textbf{p}}$ for narrow bands $\Delta k$ and $\Delta p$ at $p=k$.
Therefore, for $\textbf{k}, \textbf{p} \leq \textbf{k}^{max}/2,$ the numbers of above-condensate atoms
in the $\textbf{k}$- and $\textbf{p}$-representations are the same
($\sum\limits_{\textbf{k}}N_{\textbf{k}} \approx f(\textbf{k})N k^{2}\Delta k /(2\pi^{2})$).

However,   the solution for the wave function
$\Psi_{0}$ under the zero BCs \cite{zero-liquid,zero-gas1} differs from that under the cyclic BCs.
In view of this, the values of $\rho_{\infty}$ must also be different. Hence, the amount of the condensate $\rho_{\infty}N$ and the number of
above-condensate atoms
under the zero BCs must \textit{differ} from those under the cyclic BCs,  in spite of the commonly accepted expectations.
 But we have not calculated $\rho_{\infty}$ in the present work.

\section{Discussion}

It is usually considered that the boundaries have no influence
on the bulk properties of a system, because the near-boundary region is small
as compared with the total volume. However, the result obtained in the article is topological and remains valid at an arbitrarily thin
near-boundary region. The boundaries should exist as a region, where the properties of the system are sharply changed.

We mention one more significant item. At cyclic boundaries,
we must expand the density matrix in cyclic eigenfunctions. But, in the presence of
boundaries, we may expand in noncyclic eigenfunctions, e.g., in sines with components
of $k$ of the form $n\pi/L$, as it was made in the article.
Such an expansion is not quite proper, in our opinion, for a cyclic system,
since such functions separate two points, namely the edges of the system; but
a cyclic system has no edges. Whereas, the edges do exist in a system with boundaries;
therefore, the presence of boundaries allows one to use
$\textbf{k}$-representation. It is an additional language. Apparently, it is not only the language:
the boundaries  reveal some microstructure of a system, which is absent without boundaries.

Is it possible to obtain an experimental confirmation? We can
propose only rough preliminary considerations. If we represent the
ordinary means of detecting of a condensate (by neutron peaks, e.g.) in terms of
eigenfunctions (\ref{5}), then the separate macroscopically filled
levels (\ref{16}) will be probably insoluble in experiments due
to very small distances between them. However, the spreading of a
condensate over levels can be established, most likely, by
comparing the theory with the peculiarities of scattering peaks.
Another means is to modify the Bogolyubov model \cite{bog1947} for
the account for boundaries, by using sines instead of exponential
functions. Due to the macroscopic population, the lowest even
levels should be identify with the condensate, rather than with
quasiparticles, so that these levels turn out forbidden for quasiparticles. This
rather strange property can be observable. For gases
in a trap, the occupation  of the lowest levels must be affected by
the trap field (it varies on scales of the order of system's size,
i.e., of the order of the wavelength of the lowest levels of the
condensate (\ref{16})). This can be calculated and be apparently
measured.


Notice that the fragmented condensate (\ref{17}) belongs to the generalized condensates,
which was considered in several works, but only for free particles
or point-like potentials or cyclic BCs (see, e.g., \cite{gir1965,ziff1977,berg1982} and the recent review  \cite{mullin2012}). In this case, the properties
of the ideal gas are strongly different from those of an interacting gas. In particular, it is impossible to obtain a fragmented condensate with the
help of the above-mentioned reasoning. Indeed, all particles of the ideal gas in the ground state are characterized by some wave function $\psi(\textbf{r})$.
Then $\rho(\textbf{r}_{1},\textbf{r}_{2})$
(\ref{2}) can be exactly calculated:
\begin{equation}
\rho(\textbf{r}_{1},\textbf{r}_{2})= V\psi^{*}(\textbf{r}_{1})\psi(\textbf{r}_{2}).
\label{34}  \end{equation}
For $|\textbf{r}_{1}-\textbf{r}_{2}|\gg \bar{R},$ the quantity $ \rho(\textbf{r}_{1},\textbf{r}_{2})$ does not approach the constant
$\rho_{\infty}$, since $\psi(\textbf{r})\neq const$. In particular, $\psi(x)=const\times\sin{(\pi x/L)}$
at a high barrier in the one-dimensional case. For such $\psi(x)$, the particle are mostly localized at the center of a vessel.
In this case, formulas (\ref{14}) and (\ref{16}) are wrong.
The ground state corresponds to the case where all particles are in the single lowest state. Hence, (\ref{10b})
yields $N_{\textbf{k}}=\delta_{\textbf{k},\textbf{k}_{1}}N,$ and the condensate is present only on the lowest level.
For the cyclic boundaries, $\psi(\textbf{r})=1/\sqrt{V}$ and $\rho(\textbf{r}_{1},\textbf{r}_{2})= 1= const$,
the eigenfunctions are exponential functions (eigenfunctions of the operator of momentum), and the expansion of $1$ in exponential functions gives
the single macroscopically filled level with $k=0$.
In other words, the condensate is present only on the single level both in a cyclic system and in a system with boundaries.
This corresponds to the results by Ziff et al. \cite{ziff1977}, according to which the bulk results of the ideal gas
are independent of the wall BCs.

The effects revealed in the article and  in
\cite{zero-liquid,zero-gas1} are related to the interatomic
interaction. For fluids and gases, namely the interaction
distributed the particles uniformly. Therefore,
$\rho(\textbf{r}_{1},\textbf{r}_{2})
|_{|\textbf{r}_{1}-\textbf{r}_{2}|\gg
\bar{R}}=\rho_{\infty}=const$, which yields the fragmented
condensate (\ref{16}). The nature of the  effect
\cite{zero-liquid,zero-gas1} is more complicated. Here, the
interatomic interaction is necessary, since the effect is related
to bulk waves (collective oscillations), their distributions
over harmonics, and, apparently, to the interaction of
harmonics.  At present, we try to check solutions  \cite{zero-liquid,zero-gas1} by another methods.

\section{Conclusion}
The analysis shows that the presence of boundaries allow one to propose the new representation of the condensate,
at which the condensate is dispersed over many lowest one-particle levels.
This effect is related to that the walls of a vessel play the role of a resonator and affect the spectrum of eigenmodes of the system.

The proposed $\textbf{k}$-representation is equivalent, on the whole, to the traditional one,
but it allows one to reveal new properties of a system, which are hidden, if the
ordinary $\textbf{p}$-representation is used.

This work is partially supported by the Special Program of Fundamental Studies
of the Division of Physics and Astronomy of the National Academy of Sciences of Ukraine.

     \renewcommand\refname{}


\begin{thebibliography}{200}
\bibitem {einstein25} A. Einstein, Sitzungsber. preuss. Akad. Wiss. \textbf{1}, 3 (1925).
\bibitem {bog1947} N.N. Bogoliubov, J. Phys. USSR \textbf{11}, 23 (1947).
\bibitem {penronz1956} L. Penrose and O. Onsager, Phys. Rev. \textbf{104}, 576 (1956).
\bibitem {blag}  I.V. Bogoyavlenskii, L.V. Karnatsevich, Zh.A. Kozlov, and A.V. Puchkov, Sov. J. Low Temp. Phys. \textbf{16}, 140 (1990).
\bibitem {cond2000} H.R. Glyde, R.T. Azuah, and W.G. Stirling, Phys. Rev. B \textbf{62}, 14337 (2000).
\bibitem {gfmc1981} M.H. Kalos, M.A. Lee, P.A. Whitlock, and G.V. Chester, Phys. Rev. B \textbf{24}, 115 (1981).
\bibitem {vak-cond} I.A. Vakarchuk, Theor. Math. Phys. \textbf{82}, 308 (1990).
\bibitem {reatto1994}  L. Reatto, G.L. Masserini, and S.A. Vitiello, Physica B \textbf{197}, 189 (1994).
\bibitem {cep1995}   D.M. Ceperley, Rev. Mod. Phys. \textbf{67}, 279 (1995).
\bibitem {me2006}  M. Tomchenko,  Low Temp. Phys. \textbf{32}, 38 (2006).
\bibitem {rota2012}  R. Rota, J. Boronat, J. Low Temp. Phys. \textbf{166}, 21 (2012).
\bibitem {gir1965}   M.D. Girardeau, J. Math.  Phys. \textbf{6}, 1083 (1965).
\bibitem {zero-liquid}  M.D. Tomchenko, Ukr. J.~Phys. \textbf{59}, 123 (2014); arXiv:cond-mat/1201.1845.
 \bibitem {land3} L.D.~Landau, E.M.~Lifshitz, {\it Quantum Mechanics.
Non-Relativistic Theory} (Pergamon, New York, 1980).
\bibitem {ziff1977} R.M. Ziff, G.E. Uhlenbeck, and M. Kac, Phys. Rep. \textbf{32C}, 169 (1977).
\bibitem {berg1982} M. van den Berg and  J.T. Lewis,  Physica A \textbf{110}, 550 (1982).
\bibitem {mullin2012}  W.J. Mullin, A.R. Sakhel, J. Low Temp. Phys. \textbf{166}, 125 (2012).
\bibitem {zero-gas1}  M. Tomchenko,   arXiv:cond-mat/1204.2149.




\end{thebibliography}
       \end{document}


\title{Erratum: Structure of a condensate for the Bose fluid in a vessel}
\author{Max Tomchenko}  \email{mtomchenko@bitp.kiev.ua}
\affiliation{Bogolyubov Institute for Theoretical Physics, 14-b Metrolohichna Street,
Kiev 03680, Ukraine}

\date{\today}
\begin{abstract}
We discuss a possibility of a fragmented structure
of the condensate for an interacting Bose systems placed in a vessel.
The erroneous result of the main part of this work is corrected.
\end{abstract}

\pacs{67.10.Ba, 67.25.D-}
\maketitle

In the main part of this archival work, we considered the structure of a condensate for the Bose fluid placed in a vessel.
Since the vessel breaks the translation invariance of the system,
the momentum is a ``bad'' quantum number. Therefore, we proposed to expand the density matrix
in sines, which are the eigenfunctions of a free particle in the box.
However, our conclusion about the possibility of the expansion in sines is erroneous at least for a fluid or gas. The error arose because we introduced
the condensate in the way used for the homogeneous systems, whereas it is necessary to use a more general approach to the inhomogeneous systems \cite{leggett2006}:
\begin{equation}
\rho(\textbf{r}_{a},\textbf{r}_{b}) = \sum\limits_{j}p_{j}f_{j}
^{*}(\textbf{r}_{a})f_{j}(\textbf{r}_{b}),
\label{1}  \end{equation}
where $f_{j}(\textbf{r})$ is some complete orthonormalized collection of functions.
We consider the time-independent ground state of the system.

The specific feature of the problem consists in that the total wave function of the system bounded by walls should be equal to zero on the boundaries.
Therefore, the one-particle density matrix $\rho(\textbf{r}_{a},\textbf{r}_{b})$
and the basis functions $f_{j}(\textbf{r})$ should also be zero on the boundaries.
Most likely, we can choose as $f_{j}(\textbf{r})$ the ordinary exponential functions $c_{\textbf{k}}e^{i\textbf{k}\textbf{r}}$
with a small additions $\delta f_{\textbf{k}}(\textbf{r})$, which are nonzero only near the boundaries.
Then the condensate will be defined in a usual way, and the presence of walls will reveal itself only as the exhaustion of the condensate near the walls.
But we do not know whether it is possible to construct such collection of functions $f_{j}(\textbf{r})$.

If not, this means that, most likely, the basis functions do not include the function $f_{0}(\textbf{r})\approx const$
that leads to the ordinary condensate. Then the condensate has an unusual structure and, apparently,
is spread over several or many levels (the so-called fragmented BEC:  several of $p_{j}$
in (\ref{1}) are of  order a number $N$ of bosons,
and the rest ones of  order unity; we assume the normalization,
at which $\rho(\textbf{r}_{a},\textbf{r}_{b})\simeq N/V$ far from the walls).

Nozieres \cite{nozieres1995} showed that a condensate is energy-gained,
being on the single level rather than on several ones.
His proof assumed that the system can be in the state with the single macroscopically filled level.
However, the expansion of the density matrix in eigenfunctions may be such that it is impossible to have the
condensate only on a single level.
In this case, a fragmented condensate is possible.

To our knowledge, the ground-state wave function of a system of interacting bosons in a box at an arbitrary nonpoint-like interatomic potential was found only in work \cite{zero-liquid}. The results in \cite{zero-liquid} imply that
$\rho(\textbf{r}_{a},\textbf{r}_{b})= const$ far from the walls,
$\rho(\textbf{r}_{a},\textbf{r}_{b})$ must oscillate at distances of the order
of the interatomic one from the walls, and $\rho(\textbf{r}_{a},\textbf{r}_{b})= 0$ on the walls.
The function $\rho(\textbf{r}_{a},\textbf{r}_{b})$ has a very complicated structure;
therefore, it is easier to solve the problem for a weak interaction. We did not study this problem yet.
However, it is worth solving, since some surprises are not excluded.
In particular, we mention the unexpected result  \cite{zero-liquid} that
the presence of walls changes strongly the ground-state energy
of the system and the dispersion law of phonons.  We cannot exclude the influence of walls on the
condensate structure as well.

    \renewcommand\refname{}